\documentclass[aip,jcp,reprint]{revtex4-1}

\usepackage[colorlinks=true,linkcolor=black,citecolor=black,urlcolor=black,bookmarks,breaklinks=true]{hyperref}
\usepackage{anyfontsize}
\usepackage{enumitem}
\usepackage{graphicx}
\usepackage{xcolor}

\usepackage{amsthm,amsmath,amssymb,amsfonts}
\usepackage{bm}

\usepackage[caption=false]{subfig}

\usepackage{booktabs}
\usepackage{multirow}
\usepackage{dcolumn}
\newcolumntype{d}[1]{D{.}{.}{#1}}

\usepackage{soul}
\setstcolor{red}

\newcommand{\R}{\mathbb{R}}

\newcommand{\E}[1]{\left\langle#1\right\rangle}
\newcommand{\vect}[1]{\mathbf{#1}}
\newcommand{\fn}[2]{\mathinner{#1\mathopen{\left(#2\right)}}}
\newcommand{\nv}[1]{\fn{\sigma_N^2}{#1}}
\begin{document}

\title{Local number fluctuations in ordered and disordered phases of
       water across temperatures: Higher-order moments and degrees of
       tetrahedrality}

\author{Michael A. Klatt}
\email[]{michael.klatt@dlr.de}
\affiliation{German Aerospace Center (DLR), Institute for AI Safety and Security, Wilhelm‐Runge‐Str.~10, 89081 Ulm, Germany}
\affiliation{\hbox{German Aerospace Center (DLR), Institute for Material Physics in Space, 51170 Köln, Germany}}
\affiliation{\hbox{Department of Physics, Ludwig-Maximilians-Universit\"at M\"unchen, Schellingstr.~4, 80799 Munich, Germany}}

\author{Jaeuk Kim}
\email[]{jaeukk@princeton.edu}
\affiliation{Department of Chemistry, Princeton University, Princeton, NJ 08544, USA}
\affiliation{Department of Physics, Princeton University, Princeton, NJ 08544, USA}
\affiliation{\hbox{Princeton Materials Institute, Princeton University, Princeton, NJ 08544, USA}}

\author{Thomas E. Gartner III}
\email[]{teg323@lehigh.edu}
\affiliation{\hbox{Department of Chemical \& Biomolecular Engineering, Lehigh University, Bethlehem, PA 18015, USA}}

\author{Salvatore Torquato}
\email[]{torquato@princeton.edu}
\affiliation{Department of Chemistry, Princeton University, Princeton, NJ 08544, USA}
\affiliation{Department of Physics, Princeton University, Princeton, NJ 08544, USA}
\affiliation{\hbox{Princeton Materials Institute, Princeton University, Princeton, NJ 08544, USA}}
\affiliation{\hbox{Program in Applied and Computational Mathematics, Princeton University, Princeton, NJ 08544, USA}}

\date{\today}

\begin{abstract}
The isothermal compressibility (i.e., the asymptotic number variance) of
equilibrium liquid water as a function of temperature is minimal near
ambient conditions. This anomalous non-monotonic temperature dependence
is due to a balance between thermal fluctuations and the formation of
tetrahedral hydrogen-bond networks. Since tetrahedrality is a many-body
property, it will also influence the higher-order moments of density
fluctuations, including the skewness and kurtosis. To gain a more
complete picture, we examine these higher-order moments that encapsulate
many-body correlations using a recently developed, advanced platform for
local density fluctuations. We study an extensive set of simulated
phases of water across a range of temperatures (80\,K to 1600\,K) with
various degrees of tetrahedrality, including ice phases, equilibrium
liquid water, supercritical water, and disordered nonequilibrium
quenches. We find clear signatures of tetrahedrality in the higher-order
moments, including the skewness and excess kurtosis, that scale for all
cases with the degree of tetrahedrality. More importantly, this scaling
behavior leads to non-monotonic temperature dependencies in the
higher-order moments for both equilibrium and non-equilibrium phases.
Specifically, at near-ambient conditions, the higher-order moments
vanish most rapidly for large length scales, and the distribution
quickly converges to a Gaussian in our metric. However, at non-ambient
conditions, higher-order moments vanish more slowly and hence become
more relevant especially for improving information-theoretic
approximations of hydrophobic solubility. The temperature
non-monotonicity that we observe in the full distribution across
length-scales could shed light on water's nested anomalies, i.e., reveal
new links between structural, dynamic, and thermodynamic anomalies.
\end{abstract}


\maketitle

\section{Introduction}

\begin{figure}[t]
  \centering
  \includegraphics[width=\linewidth]{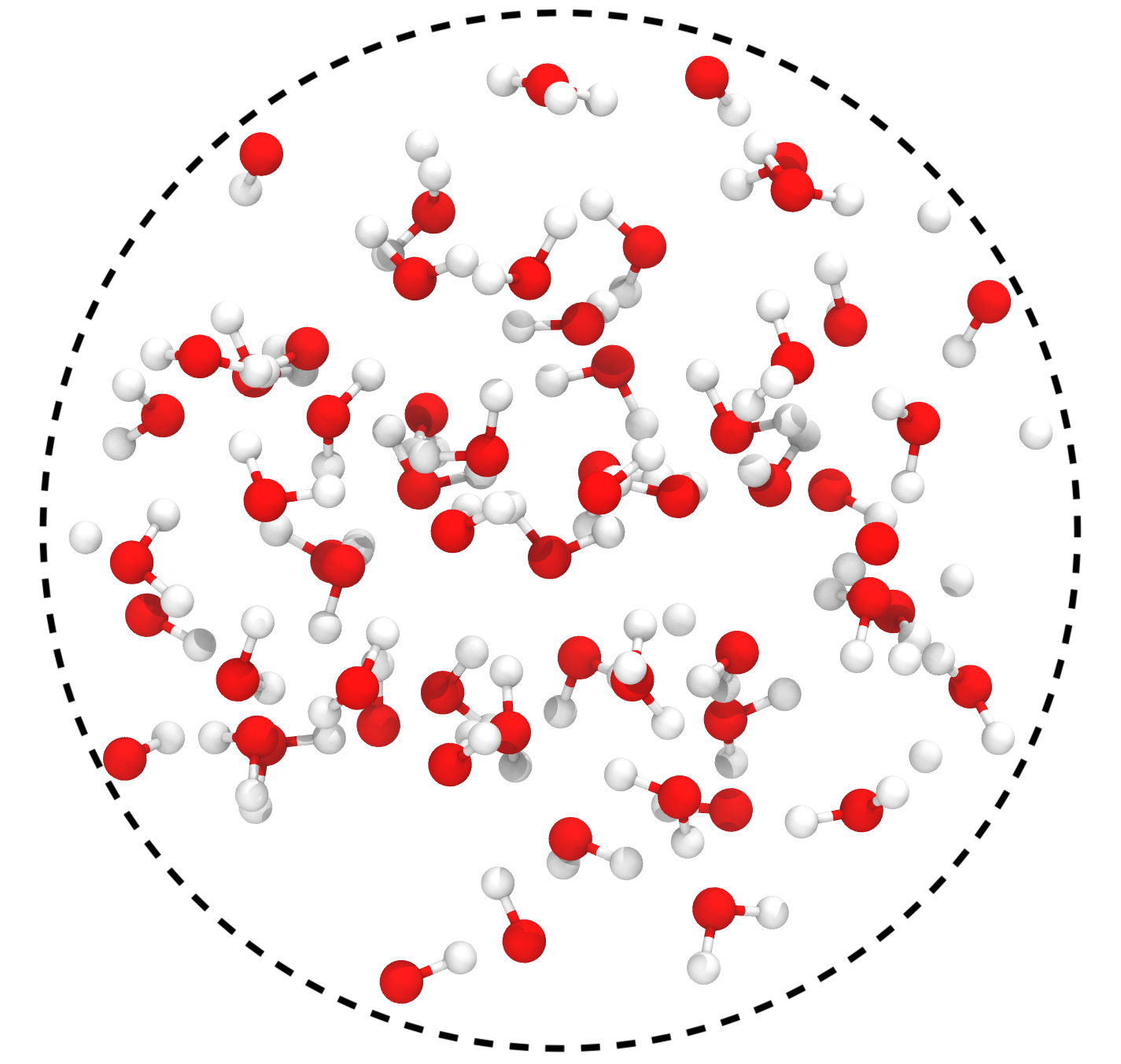}
  \caption{Snapshot of water molecules in the TIP4P/2005 model in the
  equilibrium liquid at $T = 300$\,K. Density fluctuations are evaluated
  by the number distribution of oxygen atoms within a spherical
  observation window, here with radius $R=7\,$\AA.}
  \label{fig:snapshot}
\end{figure}

Physical properties of many-particle systems can be greatly influenced
by their density fluctuations.\cite{Sc66, Ca78, Ha86, Jo91, Pe93, To00a,
La03,To08b, klatt_genuine_2022} The relationship between the isothermal
compressibility and infinite-wavelength density fluctuations in thermal
equilibrium is a classic example.\cite{Ha86, To03a} More generally, in
any many-particle system both in and out of equilibrium, local density
fluctuations can be comprehensively quantified via the probability
distribution $P[N(R)]$, where $N(R)$ is the number of particles in a
spherical observation window of
radius~$R$;\cite{torquato_local_2021,zheng_hidden_2021} see
Fig.~\ref{fig:snapshot}. The variance of this probability distribution,
i.e., the number variance~$\sigma_N^2(R)$, and its large-$R$ scaling is
a key determinant of the physical and structural properties of many-body
systems, including elastic moduli, electronic transport properties, and
electromagnetic properties.\cite{Sc66,Ha86,To03a,To18a, To21a, Vy23}
Controlling $\sigma_N^2(R)$ can lead to optimal mechanical
responses,\cite{xu_microstructure_2017} optimal transport
properties,\cite{torquato_predicting_2020, klatt_critical_2021,
klatt_wave_2022} and superior strategies for
sensing~\cite{jiao_avian_2014} or learning.\cite{monderkamp_active_2022} 

The fact that water is so abundant and of critical importance in
biological and industrial contexts has motivated many investigators to
understand and quantify relationships between its complex physical
properties and structure, especially its density
fluctuations.\cite{chandler_gaussian_1993, Hu96, garde_origin_1996,
lum_hydrophobicity_1999, pratt_molecular_2002, wolde_hydrophobic_2002,
chandler_interfaces_2005, pratt_theory_2008, lebard_dynamical_2008,
mittal_static_2008, russo_water-like_2018} Water's intricate local
structure exhibits temperature- and pressure-dependent shifts in a
generally tetrahedral hydrogen bond network, which strongly affects the
density fluctuations of water. A balance between thermal fluctuations
and the formation of such a tetrahedral network is believed to be the
reason for liquid water to anomalously exhibit extrema in thermodynamic
response functions upon cooling at constant
pressure.\cite{debenedetti_supercooled_2003} For example, the isothermal
compressibility of liquid water in equilibrium as a function of
temperature is minimal near ambient conditions, which directly
translates into a corresponding minimum of the large-$R$ asymptotic
number variance. Another example where the analysis of density
fluctuations, specifically $\nv{R}$, has proven useful to predict the
physical properties of water (mediated by its tetrahedral nature) is the
solubility of hydrophobic solutes in
water.\cite{godawat_characterizing_2009, patel_fluctuations_2010,
patel_extended_2011, patel_quantifying_2011, singh_two-state_2016,
jiang_characterizing_2019, ashbaugh_bridging_2021,
sinha_connecting_2022, rego_understanding_2022} Finally, two-body
information and tetrahedrality have also been used to study density
fluctuations in amorphous ices as well as transitions between their
different forms.\cite{martelli_large-scale_2017,
martelli_searching_2018, gartner_manifestations_2021,
formanek_molecular_2023}

Tetrahedrality is inherently a many-body property, involving at least
one molecule and its four neighbors. It, therefore, affects not only
second-order properties, like $\nv{R}$, but also higher-order moments of
$P[N(R)]$, specifically, its skewness and kurtosis. To better understand
the complex and anomalous behaviors in the physical properties of water,
there is much to be gained via a complete description of the dependence
of water's density fluctuations on its tetrahedrality. While there have
been previous efforts to understand such effects by using typical
two-body statistics,\cite{nilsson_structural_2015-1,
sellberg_ultrafast_2014} (see also Fig.~\ref{fig:G2} in
Appendix~\ref{app:G2}) these approaches may provide an incomplete
picture because they lack the effects of tetrahedrality on three- and
four-body properties. Thus, it is desired to extract information from
the higher-order moments and full distribution $P[N(R)]$ on local length
scales, i.e., to go beyond the two-body statistics contained in
$\nv{R}$.

Here, we study in detail the higher-order moments of density
fluctuations in water using a recently developed, advanced ``platform''
or ``toolset'' for local density fluctuations.\cite{torquato_local_2021}
Thus, we gain a refined physical and structural understanding of the
impact of tetrahedrality. This approach highlights the fundamental
importance of higher-order structural information to fully characterize
density fluctuations across length scales. The analysis is applicable to
generic many-particle systems and is based on explicit closed-form
integral expressions for structural information up to three- and
four-body correlations, rigorous bounds, and high-precision numerical
techniques.

We apply this new toolset to an extensive set of water states across the
phase diagram with various degrees of tetrahedrality, simulated via the
TIP4P/2005 model.\cite{abascal_general_2005} For this reason, we also
consider water states far from ambient conditions, which often have been
neglected in previous studies about water's density fluctuations.
Specifically, we consider ($i$) liquid water in equilibrium at ambient
temperature (300\,K), close to the liquid-vapor critical point
(646.4\,K), and in a supercritical state (1600\,K); see
Fig.~\ref{fig:snapshot} for a snapshot of water molecules at $T$ =
300\,K; ($ii$) supercooled liquid water at $T =$ 200\,K; ($iii$)
hexagonal ice~Ih and cubic ice~Ic just below this model's melting
temperature at $T =$ 250\,K; and ($iv$) nonequilibrium quenches of water
at $T =$ 80, 180, 190, 200\,K.
Such diverse phases of water, some far from ambient, are relevant for
applications such as biopreservation\cite{weng_exploring_2019}, climate
modeling\cite{tabazadeh_surface_2002, yang_ice_2021}, astrobiology/space
exploration\cite{mottl_water_2007}, and understanding life near
hydrothermal vents.\cite{mcdermott_pathways_2015}
We compare our states of water to two further reference systems with low
or high degrees of tetrahedral order, respectively: equilibrium hard
spheres (as a prototypical example of a simple
liquid)~\cite{hansen_theory_2013,torquato_random_2002} and a continuous
random network (as a classical model of amorphous
silicon).\cite{barkema_high-quality_2000} 

Our key findings are briefly summarized as follows:
\begin{itemize} 
  \item We show that higher-order moments contain many-body information
    and hence detect tetrahedrality, i.e., they exhibit distinct
    features that scale with the degree of tetrahedrality;
  \item While lower-order moments work best close to room temperatures,
    we demonstrate  that higher-order moments are needed at both higher
    and lower temperatures;
  \item We show that higher-order moments improve information-theoretic
    approximations of the hydrophobic solubility at far-from-ambient
    conditions.
\end{itemize}

More specifically, our results reveal how the degree of tetrahedrality
is captured by explicit features in the skewness and excess kurtosis at
specific radii $R$; see Sec.~\ref{signatures} and
Figs.~\ref{fig:pair-statistics}--\ref{fig:tetra2}. These features scale
for all of our phases with the degree of tetrahedrality and thus offer
an additional, statistically robust characterization of tetrahedral
order in disordered phases of water both in and out of equilibrium.

More importantly, in Sec.~\ref{clt}, we find that the tetrahedrality
also leads to non-monotonic temperature dependencies in the higher-order
moments and consequently a Gaussian distance metric. Whereas the
well-known minimum in the isothermal compressibility as a function of
temperature only predicts a minimal \textit{asymptotic} variance, we
here observe a non-monotonic behavior also in the higher-order moments
in local density fluctuations. For intermediate to large radii, the
distance to a Gaussian distribution vanishes most rapidly at
near-ambient conditions. At both higher and lower temperatures, the
decay slows down; see Fig.~\ref{fig:l2}. Hence, higher-order moments are
more relevant at non-ambient than at ambient conditions. As discussed
below, we surmise that this non-monotonicity again results from a
balance between tetrahedrality and thermal fluctuations, but in our case
this trade-off affects the higher-order moments. The observed tendency
of slower convergence cannot be predicted solely from the number
variance or other two-body information.\cite{torquato_local_2021}

Consequently, the accuracy of any second-order approximation that
assumes Gaussian density fluctuations will deteriorate as the
temperature moves away from ambient conditions. As a prominent example,
we study this effect for the information-theoretic approximations of
hydrophobic solubility;\cite{Hu96} see Sec.~\ref{ITA}. Previous studies
at ambient conditions found that higher-order moments do not improve
upon the second-order approximation.\cite{Hu96,hummer_hydrophobic_1998}
While we confirm this finding for equilibrium water at $T=300$\,K, we
show that for all other state points we consider, higher-order moments
indeed improve the approximation, i.e., equilibrium water around room
temperature is again an exception. More specifically, we find that the
approximation is accurate only for a range of radii where the number of
constrained moments essentially determines the entire probability
distribution; the higher the number of moments, the larger this range of
radii; see Fig.~\ref{fig:ITA_flat}. 

The structural differences revealed by our analysis of the higher-order
moments directly translate into different solvation behaviors. For
example, we show that for strongly negative values of the skewness, the
information-theoretic prediction of the hydrophobic solubility tends to
overestimate the excess chemical potential. For a review of the
far-reaching chemical implications of hydrophobic solubility; see
\citet{rego_understanding_2022}.
Moreover, our structural characterization of water at all length scales
and the non-monotonic temperature dependence that we discovered in the
higher-order moments may reveal, in future work, structural links
between the known structural, dynamic, and thermodynamic anomalies of
water\cite{errington_relationship_2001}, as briefly discussed in
Sec.~\ref{sec:conclusions}.

Even though experimentally accessible pair correlations might look
similar for two liquid-like states of water, there can be distinct
structural differences between these states. Distinguishing those states
by their higher-order moments improves the accuracy of
information-theoretic approximations of hydrophobic solubility 
far from ambient conditions. Moreover, the non-monotonic temperature
dependence of the higher-order moments could shed light on water's
nested anomalies (structural, dynamic, and
thermodynamic).\cite{errington_relationship_2001} Thus, our study
further motivates the need for experimental methods to ascertain three-
and higher-body correlations in water systems.
\cite{soper_estimating_1995, dhabal_triplet_2014, dhabal_probing_2017}

We provide details on the models, phases and structural characteristics
(together with their mathematical definitions) in
Sec.~\ref{sec:methods}, before we present our results in
\ref{sec:results} (as described above). Finally, we provide concluding
remarks in Sec.~\ref{sec:conclusions}.

\section{Methods} \label{sec:methods}

We use extensive, state-of-the-art simulations to represent a
broad range of water phases and reference states. We then analyze
these samples with the methods developed by Torquato, Kim, and
Klatt.\cite{torquato_local_2021}

\begin{table*}[t]
  \caption{
  Simulation parameters for all water states of the TIP4P/2005 model
  considered here, where $T$ is the temperature, $N$ is the number of
  oxygen atoms inside the simulation box, $N_c$ is the number of
  configurations, $L_x$, $L_y$, $L_z$  are the side lengths of the
  simulation box (where $\ast$ indicates a cubic simulation box), and
  $\rho$ is the number density.
  \label{tab:models}
  }
\centering
\renewcommand{\arraystretch}{1.2}
\begin{tabular*}{\textwidth}{@{\extracolsep\fill}l l d{4.1} c r c c c c c}
\hline
\hline
\addlinespace[0.2em]

\multicolumn{1}{l}{Phases}                               &         & \multicolumn{1}{r}{$T$ [K]}  & $N$  & $N_c$  & $L_x$ [nm]  & $L_y$ [nm]  & $L_z$ [nm]  & $\rho$ [nm$^{-3}$]  & $\rho^{-1/3}$ [nm] \\
\hline
\addlinespace[0.2em]
\multicolumn{1}{l}{\multirow{2}{*}{Solid}} & Ice Ih      & 250.0   & 35\,152                      & 500  & 10.17  & 9.57        & 11.74       & 30.8 & 0.319 \\
                                           & Ice Ic      & 250.0   & 32\,768                      & 500  & 10.21  & $\ast$      & $\ast$      & 30.8 & 0.319 \\
\hline
\addlinespace[0.2em]
\multicolumn{2}{l}{\multirow{2}{*}{Equilibrium Liquid}}  
                                                         & 200.0   & 36\,424                      & 500  & 10.29  & $\ast$      & $\ast$      & 33.4 & 0.311 \\
                                           &             & 300.0   & 36\,424                      & 500  & 10.29  & $\ast$      & $\ast$      & 33.4 & 0.311 \\
\hline
\addlinespace[0.2em]
\multicolumn{2}{l}{\multirow{4}{*}{Quench}}              
                                                         & 80.0    & 36\,424                      & 500  & 10.29  & $\ast$      & $\ast$      & 33.4 & 0.311 \\
                                           &             & 180.0   & 36\,424                      & 500  & 10.29  & $\ast$      & $\ast$      & 33.4 & 0.311 \\
                                           &             & 190.0   & 36\,424                      & 500  & 10.29  & $\ast$      & $\ast$      & 33.4 & 0.311 \\
                                           &             & 200.0   & 36\,424                      & 500  & 10.29  & $\ast$      & $\ast$      & 33.4 & 0.311 \\
\hline
\addlinespace[0.2em]
\multicolumn{2}{l}{\multirow{2}{*}{Supercritical Fluid}} 
                                                         & 646.4   & 36\,424                      & 500  & 15.20  & $\ast$      & $\ast$      & 10.4 & 0.458 \\
                                           &             & 1600.0  & 36\,424                      & 500  & 15.20  & $\ast$      & $\ast$      & 10.4 & 0.458 \\
\hline
\hline
\end{tabular*}
\end{table*}

\subsection{Models and phases}  \label{sec:models}

We performed isothermal-isochoric (NVT) molecular dynamics (MD)
simulations of water using the TIP4P/2005 potential, one of the most
widely used water models.\cite{abascal_general_2005} TIP4P/2005 is a
rigid 4-site classical water model that faithfully reproduces water's
properties and phase diagram across a broad range of
states.\cite{vega_simulating_2011} We used GROMACS
v2018.4\cite{van_der_spoel_gromacs_2005}, integrated the equations of
motion with a leap-frog algorithm with time step 2\,fs, and used a
stochastic velocity-rescaling thermostat with relaxation time 0.1\,ps
for temperature control. We enforced bond and angle constraints with a
sixth-order LINCS algorithm, and the Van der Waals and real-space
Coulomb cutoff distances were 1.2\,nm. We used a particle-mesh Ewald
scheme to treat long-range electrostatics with a Fourier grid spacing of
0.16\,nm. We prepared disordered initial configurations using the
\texttt{gmx solvate} method at the given simulation density. 

For liquid water at  $T$ = 300\,K, we equilibrated the system for 1\,ns
and collected frames for analysis every subsequent 100\,ps. For $T$ =
200\,K metastable supercooled liquid water, we equilibrated the system
for 400 ns and collected frames for analysis every subsequent 200\,ps.
For both liquid water systems the mass density was 1.0\,g/cm$^3$. We
note that while at $T$ = 200\,K, liquid water is metastable to ice~I,
due to the separation of timescales between structural equilibration and
ice nucleation in finite-size simulations, it is possible to prepare
structurally equilibrated liquid water even at deep supercoolings. For
supercritical water at $T$ = 646.4\,K (1.01 times the liquid-vapor
critical temperature of 640\,K\cite{vega_vapor-liquid_2006}) we
equilibrated the system for 100 ns and collected frames for analysis
every subsequent 100\,ps. For supercritical water at $T$ = 1600\,K we
used a smaller time step size of 1\,fs, equilibrated the system for
10\,ns and collected frames for analysis every subsequent 100\,ps. The
supercritical systems were performed at the critical density of
0.31\,g/cm$^3$.\cite{vega_vapor-liquid_2006} For ice~Ic and Ih, we
prepared proton-disordered initial configurations using the GenIce
package\cite{matsumoto_genice_2018} at densities of 0.944\,g/cm$^3$ for
ice~Ic and 0.921 g/cm$^3$ for ice~Ih.
\cite{abascal_general_2005,zaragoza_competition_2015} We equilibrated
the ice systems for 100\,ps and collected frames for analysis every
subsequent 100\,ps. The ice simulations were performed at $T$ = 250\,K,
just below their melting temperature of $T\approx252$ K.
\cite{abascal_general_2005} For the quench configurations, we followed
exactly the stepwise quench procedure given in Gartner \textit{et
al.}\cite{gartner_manifestations_2021} at a cooling rate of 10\,K/ns,
except in this work we performed the quenches at constant volume
corresponding to a mass density of 1.0\,g/cm$^3$ instead of constant
pressure. Each configuration for analysis was taken from an independent
quench simulation. System sizes for all simulations are given in
Table~\ref{tab:models}.

To compare liquid water at $T$=300 \,K to a simple liquid, we simulate
an equilibrium hard-sphere liquid with a packing fraction $\phi=31.7\%$
and a particle number $N=10^5$ via the Monte Carlo method\cite{To02a}.
We chose the value of $\phi$ to correspond to an effective packing
fraction of liquid water near ambient
conditions\cite{jeanmairet_molecular_2013, jin_understanding_2023} in
the following sense. At unit number density, the hard-sphere diameter
$D$ is equal to the smallest distance at which the pair correlation
function of two oxygen atoms reaches one, i.e., $g_2(r= D) = 1$; for a
definition of $g_2(r)$, see Sec. \ref{sec:correlations}. To compare the
ice phases to reference systems with high tetrahedrality, we consider
amorphous silicon represented by a continuous random network with
100,000 vertices from Barkema and
Mousseau.\cite{barkema_high-quality_2000}

\subsection{Structural characterization}
\label{sec:definitions}

We quantify the density fluctuation via $P[N(R)]$ that incorporates
many-body correlations $g_n$ at unit number density. This structural
analysis is complemented by a tetrahedral order parameter $q$.

\subsubsection{Probability distribution $P[N(R)]$}
\label{sec:probabdist}

We quantify density fluctuations of many-particle systems by measuring
the probability distribution function $P[N(R)]$ via the Monte Carlo
method adopted in Torquato, Kim, and Klatt.\cite{torquato_local_2021}
Specifically, at each window radius $R$, we measure the values of $N(R)$
from $N_w(R)$ randomly placed windows in every sample configuration
under periodic boundary conditions. To reduce systematic errors due to
oversampling, we choose $N_w(R)$ such that the union volume of
observation windows of radius $R$ cannot exceed 50\% of the volume $V$
of the simulation box, i.e., $ 1-\exp\left[-N_w(R) v_1 (R)/V\right] <
0.5$, as in Torquato, Kim, and Klatt,\cite{torquato_local_2021} but with
the slight improvement that we here optimize $N_w(R)$ for each radius
separately. From the determined $P[N(R)]$, we then estimate the first
four central moments associated with number variance $\nv{R}$, skewness
$\fn{\gamma_1}{R}$, and excess kurtosis $\fn{\gamma_2}{R}$, defined as
follows, respectively:
\begin{align}
  \nv{R} := & \E{\left[\fn{N}{R} - \E{\fn{N}{R}}\right]^2} \label{eq:var}, \\
  \fn{\gamma_1}{R} := & \E{\left[ \fn{N}{R} - \E{\fn{N}{R}} \right]^3} / \left[\nv{R}\right]^{3/2} \label{eq:skewness} ,\\
  \fn{\gamma_2}{R} := & \E{\left[\fn{N}{R} - \E{\fn{N}{R}} \right]^4} / \left[ \nv{R} \right]^2 - 3 \label{eq:kurtosis},
\end{align}
where $\E{\cdot}$ represents an ensemble average. The skewness
$\fn{\gamma_1}{R}\in(-\infty,\infty)$ measures the asymmetry of the
probability distribution $P[N(R)]$, i.e., a positive value of the
skewness $\fn{\gamma_1}{R}$ means that the right tail is heavier than
the left tail, and vice versa, a negative $\fn{\gamma_1}{R}$ implies a
heavier left than right tail. A zero value of $\fn{\gamma_1}{R}$ means
that $P[N(R)]$ is symmetric around the mean value $\E{N(R)}$. The excess
kurtosis $\fn{\gamma_2}{R}\in[-3,\infty)$ measures how heavy the tails
of $P[N(R)]$ are compared to a Gaussian distribution. Specifically, a
positive value of the excess kurtosis $\fn{\gamma_2}{R}$ means that the
tails are heavier than Gaussian, and a negative value implies that the
tails are lighter. For a Gaussian distribution, both the skewness and
excess kurtosis are identically zero. To measure the deviations of
$P[N(R)]$ from Gaussian distribution, we also compute the $l_2$ distance
metric,\cite{torquato_local_2021} defined as
\begin{align}
  \fn{l_2}{R} := 
  \left\{ \left[\nv{R}\right]^{-1/2} \sum_{n=0}^\infty \left\vert \fn{F_G}{n} - \fn{F_N}{n} \right\vert^2  \right\}^{1/2}, \label{eq:l2}
\end{align}
where $F_N(n)$ and $\fn{F_G}{n}$ are the cumulative distribution
functions of $P[N(R)]$ and the discrete Gaussian distribution with mean
value $\E{N(R)}$ and variance $\nv{R}$, respectively.

\subsubsection{Many-body correlation functions $g_n$}
\label{sec:correlations}

Another route to characterize density fluctuations of many-particle
systems in $\R^d$ is to use the $n$-body correlation functions
$\fn{g_n}{\vect{r}^n}$ [or, $n$-particle probability density functions
$\fn{\rho_n}{\vect{r}^n}$] for $n\geq 2$ where ${\bf r}^n$ is a
shorthand notation\cite{Ha86,To02a} for the position vectors of any $n$
points, i.e., ${\bf r}^n:= {\bf r}_1,{\bf r}_2,\ldots,{\bf r}_n$. The
quantity $\rho_n({\bf r}^n)d{\bf r}^n$ is proportional to the
probability of finding {\it any} $n$ particles with configuration ${\bf
r}^n$ in volume element $d{\bf r}^n:= d{\bf r}_{1} d{\bf r}_{2} \cdots
d{\bf r}_n$ (see Chiu \textit{et al.}\cite{Chi13} for a mathematical
definition). When the systems are statistically homogeneous,
$\rho_n({\bf r}^n)$ is translationally invariant and hence depends only
on the relative displacements, say with respect to ${\bf r}_1$:
$\rho_{n}({\bf r}^n)=g_{n}({\bf r}_{12},{\bf r}_{13},\ldots,{\bf r}_{1n})$,
where ${\bf r}_{ij}={\bf r}_j-{\bf r}_i$. In particular, the
one-particle function is identical to the constant {\it number density}
$\rho$, and thus it is convenient to define the $n$-body correlation
function 
\begin{equation}
g_n({\bf r}^n) = {\rho_n({\bf r}^n)}/{\rho^n}.
\label{nbody}
\end{equation}
In systems without long-range order, $g_n({\bf r}^n) \rightarrow 1$ when
the points ${\bf r}^n$ are mutually far from one another. Thus, the
deviation of $g_n$ from unity measures the degree of spatial correlation
between the particles.
As in Torquato, Kim, and Klatt,\cite{torquato_local_2021} we here
compare all phases at unit number density or, equivalently, rescale them
by a length scale $\rho^{-1/3}$, which corresponds for most state points
to about 3\,\AA{} (except for the supercritical fluids, where it is
about 5\,\AA{}); see Table~\ref{tab:models}.

The important two-body function $g_2({\bf r}_{12})$ is usually called
the {\it pair correlation function}.
The {\it total correlation function} $h({\bf r}_{12})$ is defined as
$ h({\bf r}_{12})=g_2({\bf r}_{12})-1, $
and thus is a function that vanishes in the absence of spatial
correlations in the system. The structure factor $\fn{S}{\vect{k}}$ is
related to the Fourier transform of $h(\bf r)$, denoted by ${\tilde
h}({\bf k})$, via the expression
\begin{equation}
S({\bf k})=1+\rho {\tilde h}({\bf k}),
\label{factor}
\end{equation}
which is directly measurable via scattering intensity.\cite{Ki05}

\begin{figure*}[t]
  \centering
  \includegraphics[width=\textwidth]{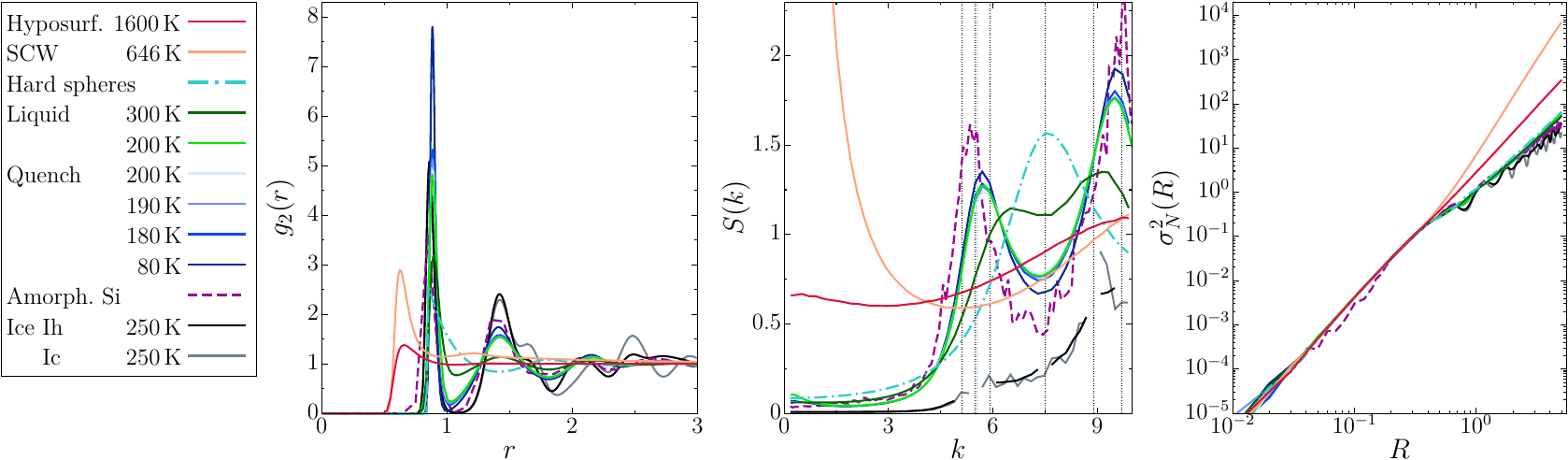}
  \caption{Two-body statistics for all states considered here:
  (a) pair correlation function~$g_2(r)$, (b) structure factor~$S(k)$,
  (c) number variance~$\sigma_N^2(R)$.
  In the legend, the states are sorted such that their mean values of
  tetrahedrality $\E{q}$ increase from top to bottom.
  As noted in Sec. \ref{sec:definitions}, all quantities are
  estimated at unit number density $\rho=1$.
  }
  \label{fig:pair-statistics}
\end{figure*}

We note that the full distribution $P[N(R)]$ and its associated moments
$\nv{R}$, $\fn{\gamma_1}{R}$ are directly related to the many-body
correlation functions $g_n$ for $n\geq 2$, as shown in Torquato, Kim,
and Klatt.\cite{torquato_local_2021} For example, the number variance
$\nv{R}$ can be computed via the two-body correlation function:
\begin{align}
  \nv{R} = \E{N(R)} \left[ 1+ \rho \int_{\R^d} h(\vect{r})\alpha_2(r;R) d{\bf r} \right],
\end{align}
where $\alpha_2(r;R)$ is the intersection volume of two spherical
windows of radius $R$, scaled by the sphere volume $v_1(R)$, whose
centers are separated by the distance $r$. Similarly, $\fn{\gamma_1}{R}$
is related to $g_2$ and $g_3$, and $\fn{\gamma_2}{R}$ is related to
$g_2$, $g_3$, and $g_4$; see Appendix \ref{app:moments-gn}.

The probability distribution $P[N(R)=m]$ can be
expressed as a series expansion involving $g_n$ for
$n=2,3,\ldots$\cite{Ve75,Zi77}. Truncating such series at the two- and
three-body levels yields sharp bounds on $P[N(R)=m]$. In particular, we
obtained two- and three-body bounds on the void probability
$P[N(R)=0]$, respectively, as follows\cite{torquato_local_2021}:
\begin{align}
  &P[N(R)=0] \leq \left[ 1- \E{\fn{N}{R}} \right] \left[ 1- \frac{\E{\fn{N}{R}}}{2} \right] + \frac{\nv{R}}{2} \\
  &P[N(R)=0] \geq \left[ 1- \E{\fn{N}{R}} \right] \left[ 1- \frac{\E{\fn{N}{R}}}{2} \right] 
  \nonumber \\
  &\qquad
  \times\left[ 1- \frac{\E{\fn{N}{R}}}{3} \right]
  + \left[ 1 -  \frac{\E{\fn{N}{R}}}{2} \right] \nv{R} 
  \nonumber \\
  &\qquad
  -  \frac{\sigma^3(R)}{6} \fn{\gamma_1}{R}.
\end{align}

\subsubsection{Tetrahedral order parameter $q$}
\label{sec:tetrahedral}

We characterize the degree of tetrahedrality of each of our state points
by the tetrahedral order parameter $q$ from Errington and
Debenedetti\cite{errington_relationship_2001} that was motivated by Chau
and Hardwick.\cite{chau_new_1998} This quantity is defined as 
\begin{align} \label{eq:q}
  q := 1 - \frac{3}{8} \sum_{j=1}^3 \sum_{k=j+1}^4 \left( \frac{1}{3} + \cos\psi_{jk} \right)^2,
\end{align}
where $\psi_{jk}$ is the angle formed by the lines joining the oxygen
atom of a given molecule to those of its (nearest) neighbors $j$ and $k$. 
Equation \eqref{eq:q} is rescaled so that its average $\E{q}$ is
0 for an ideal gas without any orientational order and 1 for a
perfect tetrahedral network.

\begin{figure}[b]
  \centering
  \includegraphics[width=\linewidth]{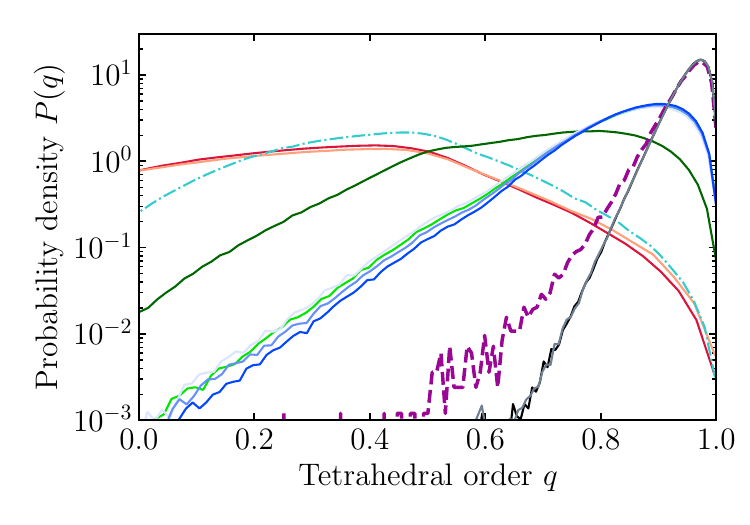}
  \caption{
  Semi-log plot of the probability density $\fn{P}{q}$ of the tetrahedral
  order parameter $q$ for all states considered here; for the legend,
  see Fig.~\ref{fig:pair-statistics}.
  }
  \label{fig:tetra_order}
\end{figure}

%%%%%%%%%%%%%%%%%%%%%%%%%%%%%%%%%%%%%%%%%%%%%%%%%%%%%%%%%%%%%%%%%%%%
%%%%%%%%%%%%%%%%%%%%%%%%%%%%%%%%%%%%%%%%%%%%%%%%%%%%%%%%%%%%%%%%%%%%

\begin{figure*}[t]
  \centering
  \includegraphics[width=\textwidth]{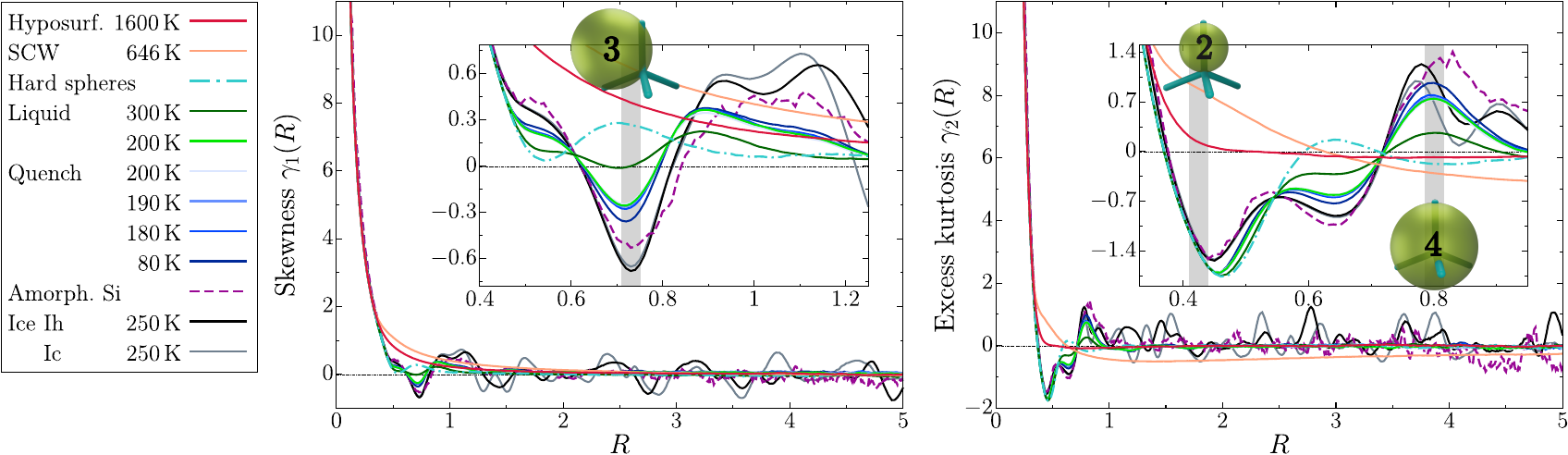}
  \caption{
  Skewness $\gamma_1$ (left) and excess kurtosis $\gamma_2$ (right)
  associated with window radius $R$ are compared for all water
  states considered here, amorphous silicon
  with a high degree of tetrahedral order, and a hard-sphere liquid with
  a low degree of tetrahedral order.
  At three specific radii (shown in the gray shades), the features in
  $\gamma_1$ and $\gamma_2$ consistently scale with the degree of
  tetrahedral order in these states.
  Correspondingly, the gray bands indicate radii where the possible
  number of atoms per observation window increases by one for a regular
  tetrahedral node. The value is indicated in the attached figures of
  the tetrahedral nodes and spherical observation windows.
  }
  \label{fig:cumulants}
\end{figure*}

%%%%%%%%%%%%%%%%%%%%%%%%%%%%%%%%%%%%%%%%%%%%%%%%%%%%%%%%%%%%%%%%%%%%
%%%%%%%%%%%%%%%%%%%%%%%%%%%%%%%%%%%%%%%%%%%%%%%%%%%%%%%%%%%%%%%%%%%%

\section{Results}
\label{sec:results}

We begin with comparisons of classical two-body statistics and
distributions of $q$ for all of our states. Strong differences in
$\E{q}$ lead to only a few distinguishing features in the two-body
statistics. Then, we turn to the higher-order moments with clear
features of tetrahedrality that point towards a non-monotonic
temperature dependence of the speed of convergence to a Gaussian number
distribution. This non-monotonicity is confirmed via $\fn{l_2}{R}$. Its
direct physical implications are discussed, among others, in the
exemplary case of hydrophobic solubility.

Figure~\ref{fig:pair-statistics} shows the two-body statistics for all
states considered here, i.e., the pair correlation function~$g_2(r)$ in
real space, the corresponding structure factor~$S(k)$ in Fourier space,
and the number variance~$\sigma_N^2(R)$ that directly quantifies density
fluctuations at the two-body level. The pair correlation function of
equilibrium water at room temperature compared to those of the
nonequilibrium quenches (down to 180\,K) differ primarily in the height
of the first two peaks and the depth of the first minimum (as a function
of the radial distance $r$), indicating the development of more
structured local environments as the temperature decreases. The ice
phases clearly exhibit Bragg peaks in $S(k)$ (dotted lines), but at
large scales, their number variance~$\sigma_N^2(R)$ has qualitatively
the same scaling as the other non-hyperuniform phases. At the critical
point, the state of water becomes anti-hyperuniform, i.e.,
$\sigma_N^2(R)$ grows faster than the volume, and $S(k)$ diverges at the
origin. In contrast, at high temperatures (above 1600\,K), water becomes
hyposurfical, i.e., the surface term in the scaling of the number
variance $\sigma_N^2(R)$ vanishes,\cite{To18a} i.e., $\sigma_N^2(R)$
exhibits a volume-like scaling even for local fluctuations.

Figure~\ref{fig:tetra_order} compares the probability density
$\fn{P}{q}$ of the tetrahedral order parameter $q$ for all of our
states. As expected, the ice phases and the amorphous silicon have the
highest degree of tetrahedrality, followed by the quenches (the higher
the final temperature, the lower the degree of tetrahedral order) and
finally the equilibrium water at $T=300$\,K. The high-temperature phases
exhibit no explicit tetrahedral order, neither does the hard-sphere
liquid.

\subsection{Signatures of tetrahedrality in higher-order moments}
\label{signatures}

We begin our analysis of the higher-order density fluctuations by
computing the third- and fourth-order central moments $\fn{\gamma_1}{R}$
and $\fn{\gamma_2}{R}$ of the number distribution, which embody up to
four-body correlations; see Eqs.~(22) and (23) in Torquato, Kim, and
Klatt.\cite{torquato_local_2021} Thus, our analysis reveals non-trivial
higher-order correlations in all states of water, including explicit
features of tetrahedrality.

Figure~\ref{fig:cumulants} shows the skewness $\gamma_1(R)$ and excess
kurtosis $\gamma_2(R)$ of the number distributions of oxygen atoms in a
spherical observation window of radius $R$.
At specific radii, $\gamma_1(R)$ and $\gamma_2(R)$ exhibit salient
features that clearly scale with the degree of tetrahedral order for all
of our states.

\begin{figure*}[t]
  \centering
  \includegraphics[width=\textwidth]{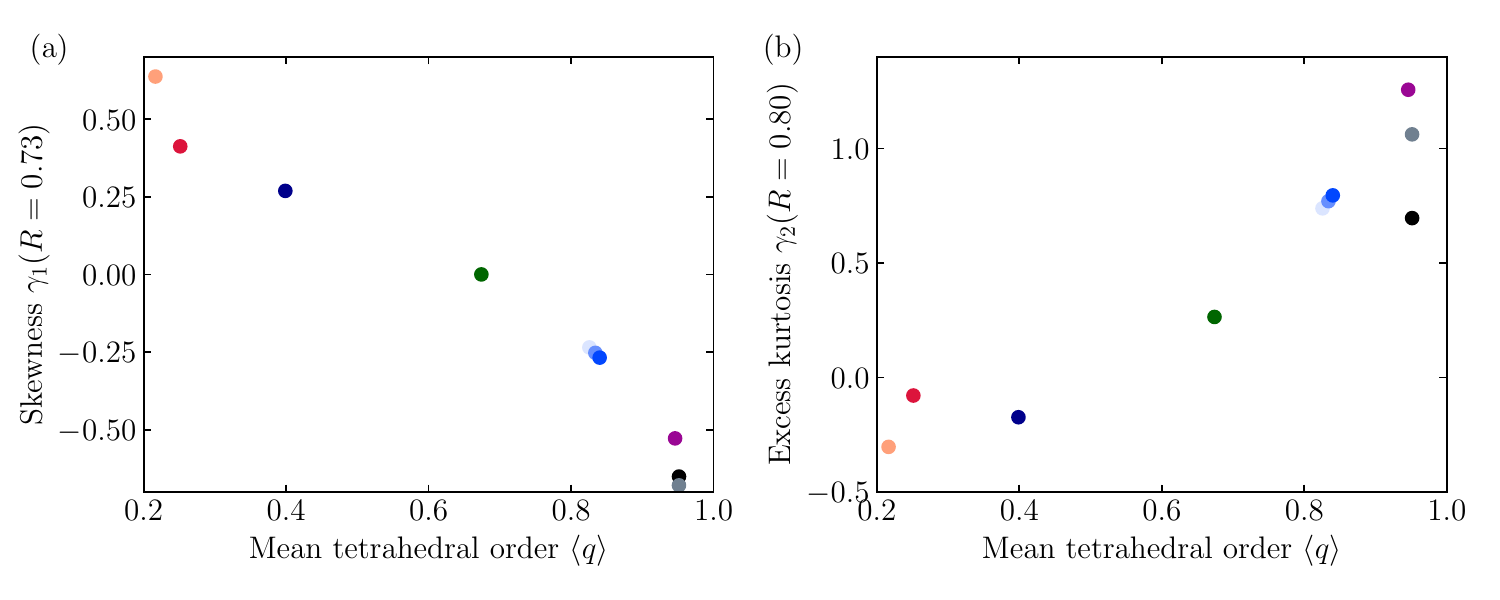}
  \caption{For all of our states, the mean tetrahedral order $\E{q}$ is
  plotted against the (a) skewness and (b) excess kurtosis at
  characteristic radii. The states are represented by the same color
  code as in Fig.~\ref{fig:cumulants}.}
  \label{fig:tetra2}
\end{figure*}

Some of these radii directly correspond to characteristic distances in a
regular tetrahedral network. To explain this relation, we first have to
choose an appropriate bond length of the tetrahedral network, e.g., here
we choose the distance to the first peak of the pair correlation
function; see Fig.~\ref{fig:pair-statistics}. Then, the gray band at
$R\approx 0.73$ in the plot of $\gamma_1(R)$ corresponds to the smallest
radius for which the spherical observation window can contain three
atoms. Close to this radius, we can therefore expect that the higher the
degree of tetrahedrality, the lower the probability of finding four or
five atoms within this radius, and hence the more left-tailed $P[N(R)]$
will be, i.e., the more negative will be the value of
$\fn{\gamma_1}{R=0.73}$.
This scaling of $\fn{\gamma_1}{R=0.73}$ with the degree of
tetrahedrality is quantitatively confirmed in Fig.~\ref{fig:tetra2}~(a),
which plots $\fn{\gamma_1}{R=0.73}$ against the average tetrahedral
order parameter~$\E{q}$.  

Attached to the gray band at $R\approx 0.73$ in the inset of
Fig.~\ref{fig:cumulants} is a small figure that indicates the
tetrahedral node and the spherical observation window, which contains up
to three atoms of the network. As we have seen, this borderline case for
three atoms imprints a clear feature of tetrahedrality on the
skewness~$\gamma_1(R)$, i.e., the third-order moment of the density
fluctuations.

The excess kurtosis~$\gamma_2(R)$, i.e., the fourth-order moment,
exhibits a similar feature at the smallest radius for which the
observation window can contain four atoms of the regular tetrahedral
network. This radius is $R\approx 0.80$, and the scaling is confirmed in
Fig.~\ref{fig:tetra2}~(b). The values for the ice phases are slightly
off, which can be explained via Fig.~\ref{fig:cumulants}. Close to
$R\approx 0.80$, $\gamma_2(R)$ exhibits a local maximum as a function of
$R$ for all of our states except at high temperatures. However, the peak
positions of the ice phases are slightly shifted with respect to those
of the disordered water phases. The latter agrees with the peak of
amorphous silicon. The different peak positions for the ice phases are
indicative of different distortions in the tetrahedral networks compared
to a regular network (e.g., different ring statistics and consequently
bending angles).

At $R= 0.42$, i.e., the smallest possible radius of a window with two
atoms of our regular tetrahedral network, we find that $\gamma_1(R)$ and
$\gamma_2(R)$ are still dominated by the large void probability, i.e.,
the probability that the window contains no atoms. Hence features that
scale with the degree of tetrahedrality only emerge at slightly larger
radii of about 0.45--0.50.

Together, these features of $\gamma_1(R)$ and $\gamma_2(R)$ offer a
refined and statistically robust characterization of tetrahedral order
in disordered phases of water both in and out of equilibrium~---~with
immediate physical implications. In fact, the additional structural
information reveals behavior in the higher-order moments that is
non-monotonic with respect to temperature.

In the search for a physical explanation of this non-monotonic behavior,
we compare $\gamma_1(R)$ and $\gamma_2(R)$ of our water states to those
of a simple liquid without tetrahedral order, namely the hard-sphere
liquid. We first focus on $R\approx 0.73$, where $\gamma_1(R)$ exhibits
the feature of tetrahedrality discussed above, i.e., the local minimum
as a function of $R$ that becomes more negative for more tetrahedrally
ordered water phases. In contrast, $\gamma_1(R)$ of the hard-sphere
liquid has a local maximum close to this radius. Its functional value,
however, is still smaller than those of the high-temperature water
states that are dominated by thermal fluctuations.

Hence, we observe that the stronger the thermal fluctuations are, the
larger is $\fn{\gamma_1}{R=0.73}$; but the stronger the tetrahedral
order is, the smaller is $\fn{\gamma_1}{R=0.73}$. At room temperature
these two influences approximately cancel out so that
$\fn{\gamma_1}{R=0.73} \approx 0$. Similar effects can be observed for
larger radii and for $\gamma_2(R)$. Overall, $\gamma_1(R)$ and
$\gamma_2(R)$ appear to converge to zero more quickly for water at room
temperature than for our other state points.

Importantly, these temperature non-monotonicities that we have found in
the higher-order moments complement the known non-monotonic temperature
dependencies of thermodynamic response functions (isothermal
compressibility and heat capacity), which anomalously increase upon
cooling. Interestingly, these anomalous temperature dependencies of the
thermodynamic response functions are not strongly reflected in the
corresponding pair correlation functions, which only show a mild
increase in the strength of correlations as temperature decreases (see
Fig.~\ref{fig:pair-statistics}). By contrast, our higher-order moments
reveal distinct structural signatures, like the pronounced features in
$\gamma_1(R)$ and $\gamma_2(R)$ at specific radii. The functional values
at these radii change with temperature and their absolute values are
minimal close to room temperature, hence mirroring the non-monotonic
trends in the response functions as functions of temperature.

In the next section, we use the Gaussian distance metric to quantify how
quickly the higher-order moments tend towards zero. We thus confirm that
the speed of convergence has a non-monotonic temperature dependence (as
surmised above); specifically, the convergence is faster at room
temperature than at higher or lower temperatures. Then, we will discuss
the chemical-physical implications for hydrophobic solubility and beyond.

%%%%%%%%%%%%%%%%%%%%%%%%%%%%%%%%%%%%%%%%%%%%%%%%%%%%%%%%%%%%%%%%%%%%

\subsection{Effect of tetrahedrality on Gaussian distance metric}
\label{clt}

\begin{figure}[b]
  \centering
  \includegraphics[width=\linewidth]{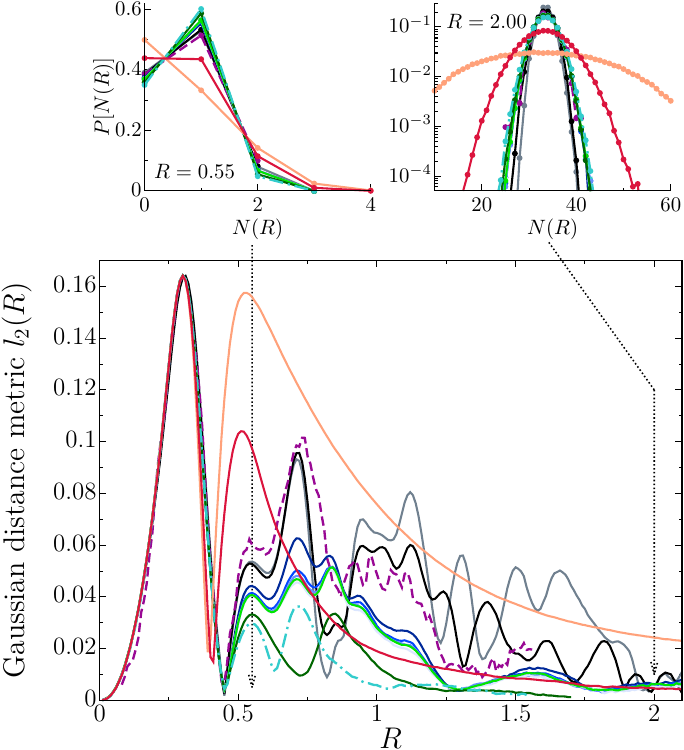}
  \caption{
  Gaussian distance metric $\fn{l_2}{R}$ as a function of window radius
  $R$; for the legend, see Fig.~\ref{fig:cumulants}. The insets show the
  number distributions at two different radii, indicated by the dotted
  arrows.}
  \label{fig:l2}
\end{figure}

The combined effect of all higher-order moments on the number
distribution can be best quantified by the Gaussian distance metric
$\fn{l_2}{R}$. Figure~\ref{fig:l2} compares our results for all states.
We observe that $\fn{l_2}{R}$ decays at large radii for all
disordered states of water (i.e., for all states except for the two ice
phases). This observation suggests that a central limit theorem (CLT)
holds for large $R$ (and thus correspondingly large $N$) with respect to
the chosen metric.

Since any CLT depends on its metric, our proposed CLT still agrees with
the heavy left tails reported for very large
radii.\cite{rego_understanding_2022} There, indirect umbrella
sampling~\cite{patel_quantifying_2011, sinha_connecting_2022} revealed
heavy left tails of $P[N(R)]$, which can be related to effects at
liquid-vapor interfaces.\cite{rego_understanding_2022} However, since
the onset of the tails appears to converge to zero in the thermodynamic
limit, they are consistent with a CLT in an $l_2$ metric.
Note that these heavy left tails in water are distinct from the more
commonly observed right tails for random point processes as reported for
a broad spectrum of models in Torquato, Kim, and
Klatt.\cite{torquato_local_2021}

The CLT and its speed of convergence for a certain range of radii
indicates how well $P[N(R)]$ can be approximated by a Gaussian
distribution and hence by two-body statistics.
As an interesting side remark, the renormalized Gaussian approximation
of Ashbaugh, Vats, and
Garde\cite{ashbaugh_bridging_2021} happens to coincide with the reference
distribution of the Gaussian distance metric
$\fn{l_2}{R}$.\cite{torquato_local_2021}

\begin{figure*}
    \includegraphics[width=\textwidth]{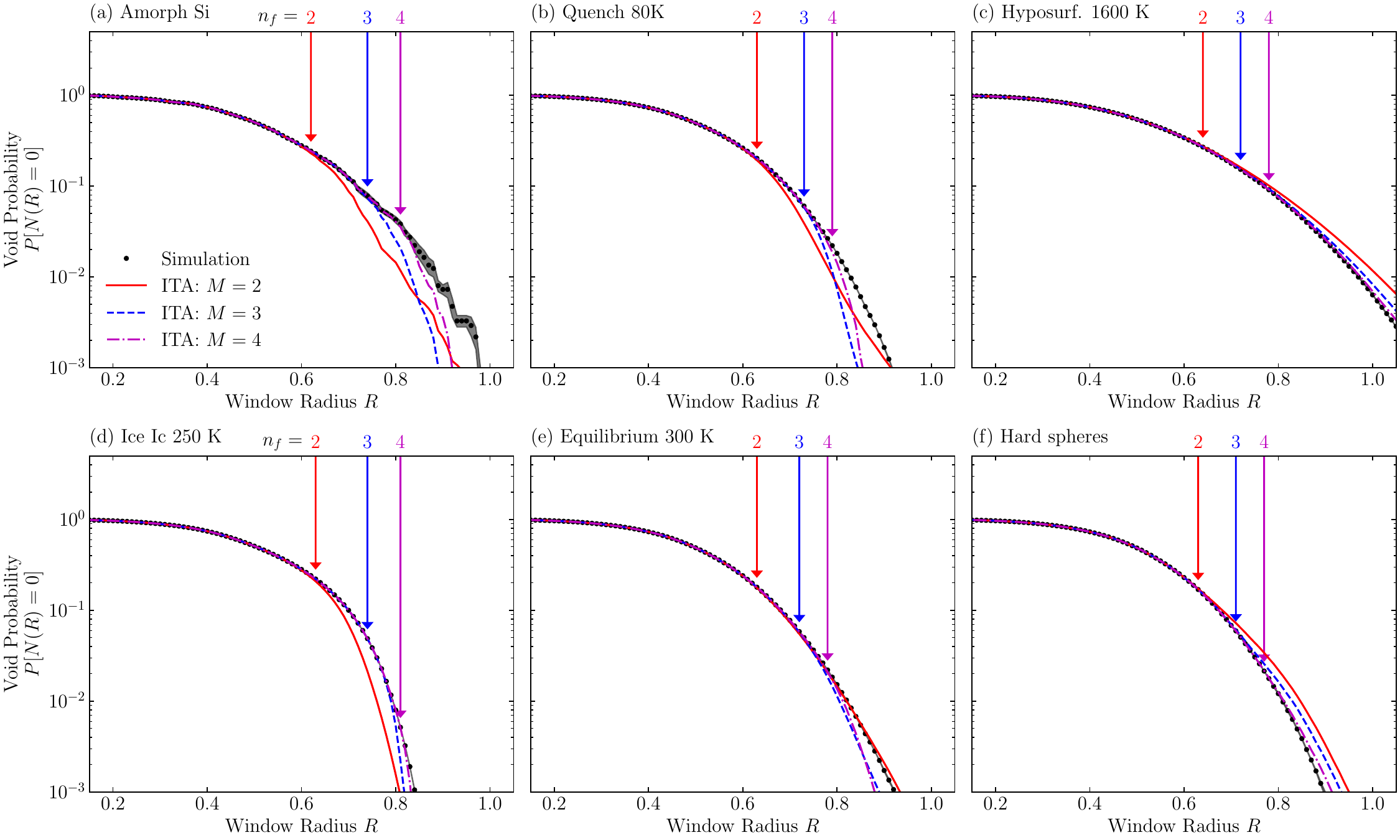}
    \caption{
    Comparison of the void probability $P[N(R)=0]$ from simulations and
    the information-theoretic approximation~---~varying the highest
    order of constrained moments $M=2,3,4$.
    The vertical arrows indicate upper bounds on the effective support
    sizes at three different values ($n_f=2,3,4$); they coincide with
    the radii at which the predictions deviate from the simulations.
    }
    \label{fig:ITA_flat}
\end{figure*}

For one relatively large radius and one relatively small radius, we also show
the corresponding number distributions in Fig.~\ref{fig:l2}. Already at
the second peak of $l_2(R)$, the Gaussian distance metric is smaller for
water at room temperature than at non-ambient conditions, whether it is
warmer or colder. This non-monotonic temperature dependence essentially
continues for larger radii and indeed confirms that among all states
considered, the number distribution converges fastest to the normal
distribution for equilibrium water at room temperature.

At high temperatures, thermal fluctuations enhance density fluctuations,
which leads to positively skewed number distributions at a given length
scale $R$; while at low temperatures, the emergence of tetrahedral order
increases the correlation length and leads to negatively skewed number
distributions. At ambient conditions, those two trends balance out, which
results in a fast convergence of the number distribution to a Gaussian
distribution.

As noted above, this balance between thermal fluctuations and
tetrahedral order is similar to the surmised reason for the anomalous
minimum in the isothermal compressibility upon cooling. In our case,
however, the non-monotonic temperature dependence appears already at a
local scale, applies to non-equilibrium phases, and it explicitly
affects the higher-order moments with immediate physical implications,
e.g., on the hydrophobic solubility, as discussed in the next section.
We will broaden the scope of the discussion in
Sec.~\ref{sec:conclusions}.

\subsection{Implications on hydrophobic solubility}
\label{ITA}

A prominent example of the chemical-physical importance of density
fluctuations in water is the solubility of hydrophobic solutes in water.
Specifically, the excess chemical potential $\mu^{\rm{ex}}$ of hydration
for a hard-particle solute of radius $R$ can be expressed by the
probability of the random formation of a spherical cavity in water,
i.e., the {\it void probability} $P[N(R)=0]$:\cite{widom_topics_1963,
beck_potential_2006}
\begin{align}
  \mu^{\rm{ex}} = -k_B T\, \ln P[N(R)=0].
  \label{eq:muex}
\end{align}
The estimation of $P[N(R)=0]$ for liquid water has been a particularly
productive field of research starting with a seminal paper by
Hummer \textit{et al.}\cite{Hu96} and subsequent works over the last two
decades.\cite{godawat_characterizing_2009, patel_fluctuations_2010,
patel_extended_2011, patel_quantifying_2011, singh_two-state_2016,
jiang_characterizing_2019, ashbaugh_bridging_2021,
sinha_connecting_2022, rego_understanding_2022}

Hummer et al.~\cite{Hu96} estimated $P[N(R)=0]$ via an
information-theoretic approximation, i.e., the probability distribution
$\hat{P}[N(R)]$ that maximizes the cross entropy (relative to a flat
prior distribution) given the first $M$ moments of $P[N(R)]$. Using the
first two moments of $P[N(R)]$, Hummer et al.~obtained (at ambient
temperatures) a surprisingly good approximation of the void probability
for small to intermediate radii, i.e., up to about
10\,\AA.\cite{rego_understanding_2022} Counterintuitively, the accuracy
of the information-theoretic approximation was found to deteriorate if
higher-order moments were taken into account, i.e., $2 < M < 7$; see Hummer
\textit{et al.}\cite{hummer_hydrophobic_1998}

This surprising accuracy at near-ambient conditions of the
information-theoretic approximation based on only the first two moments
evokes at least three questions:
\begin{itemize}
  \item[($i$)]
    Why does the information-theoretic approximation, counter-intuitively, worsen when higher-order
    moments are included?
  \item[($ii$)]
    Does the information-theoretic approximation apply to water in out-of-equilibrium or metastable
    sates, or solid phases? 
  \item[($iii$)]
    Should the tetrahedrality of water impact the accuracy of the
    prediction of the solubility?
\end{itemize}

To address these three questions, we compute (from our simulations) the
void probability $P[N(R)=0]$, also called cavity formation probability.
We then determine the information-theoretic approximation with varying
numbers of constraints $M$ (always employing the flat prior probability
as in Hummer \textit{et al.}\cite{Hu96}). Finally,
Fig.~\ref{fig:ITA_flat} compares our simulations results for $P[N(R)=0]$
(dots) to the information-theoretic approximation (colored lines).

While we confirm for water at ambient temperatures the superiority of
the information-theoretic approximation with $M=2$ over those with
higher-order moments, we find that it is an exception.
For all other states, the information-theoretic approximation is
accurate only for a range of radii where the number of constrained
moments essentially fix the entire probability distribution; the higher
the number of moments, the larger this range of radii, which can be
quantified by the following rule of thumb.

The information-theoretic approximation with $M$ constraints is accurate
if there are at most $M$ non-negligible values of $P[N(R)]$. In that
case, the $M$ moments virtually specify all probabilities. More
specifically, we define the effective support size $n_f(R)$ as the
minimal number $n$ for which $\sum_{m=n+1}^\infty P[N(R)=m]<P[N(R)=0]$.
Then, the information-theoretic approximation with $M$ constraints is
accurate for a range of radii where $n_f(R) \leq M$; see
Fig.~\ref{fig:ITA_flat}.

For the same range of radii, we also obtain accurate predictions via the
rigorous bounds from Torquato, Kim, and Klatt.\cite{torquato_local_2021}
The upper and lower bounds virtually coincide within this regime, which
may facilitate an analytic approach to hydrophobic solubility for small
radii. For larger radii, the information-theoretic approximation
prediction is more precise even though it is not perfect.

At these radii slightly larger than the constrained range, the
deviations of the information-theoretic approximation from our
simulation data display a consistent pattern.
The information-theoretic approximation underestimates $P[N(R)=0]$ when
the skewness is negative, which corresponds to a tetrahedral order
higher than that for water at ambient temperatures; and the
information-theoretic approximation overestimates $P[N(R)=0]$ when the
skewness is positive, which corresponds to a low tetrahedral order.
For water at room temperature, the skewness and excess kurtosis almost
vanish, as discussed above, resulting in accurate Gaussian predictions
of the information-theoretic approximation with $M=2$.
Hence, the surprisingly good performance of the information-theoretic
approximation at ambient conditions appears to be again related, via
$\gamma_1(R)$, to a balance of thermal fluctuations and tetrahedral
order. 

In answer to the questions above, ($ii$) the information-theoretic
approximation is a valuable approximation for all of our states, but a
similar accuracy as for water at room temperature is obtained only for
small radii, where the first $M$ moments fix the distribution. ($iii$)
Both tetrahedrality and thermal fluctuations impact the accuracy of the
information-theoretic approximation but with opposing effects, i.e.,
underestimation or overestimation of $P[N(R)=0]$, respectively. ($i$) At
room temperature, these effects roughly cancel out. For all of our other
states, higher-order moments improve the information-theoretic
approximation; more precisely, the range of radii for which we obtain
accurate predictions increases. For example, if we constrain four
instead of two moments, the range of radii with accurate predictions
increases from about $0.6$ to 0.8.

%%%%%%%%%%%%%%%%%%%%%%%%%%%%%%%%%%%%%%%%%%%%%%%%%%%%%%%%%%%%%%%%%%%%

\section{Discussion}  \label{sec:conclusions}

We have studied here the link between physical properties and density
fluctuations in water by going beyond the two-body level to higher-order
moments and by comparing a great variety of states of water across a
broad range of temperatures  80\,K to 1600\,K. These states include ice
phases, equilibrium liquid water, supercritical water, and disordered
nonequilibrium quenches, and we compared them to two further reference
systems: equilibrium hard spheres, representing a simple liquid, and a
continuous random network, representing amorphous silicon. We analyzed
all of our samples with a recently developed, advanced platform for
local density fluctuations.\cite{torquato_local_2021} This approach
includes robust estimates of higher-order moments (which enables us to
capture crucial information about $n$-body correlations), a Gaussian
distance metric.

Our analysis reveals how water's tetrahedral order affects not only the
number variance but also the higher-order moments of local density
fluctuations since tetrahedrality is a many-body property. Specifically,
we observe that the third- and four-order central moments, $\gamma_1(R)$
and $\gamma_2(R)$, scale with the mean tetrahedral order parameter
$\E{q}$ at two characteristic length scales $R=0.73$ and $0.80$,
respectively. This scaling clearly indicates that the skewness and
excess kurtosis entail signatures of tetrahedrality on the higher-order
correlations for all of our phases. Moreover, the corresponding radii
can be directly related to characteristic distances in tetrahedral
nodes.

The Gaussian distance metric $\fn{l_2}{R}$ as a function of the radius
$R$ has a local minimum at $R\approx 0.72$ for liquid water at
$T=300\,K$ but a local maximum for the quenches and the supercooled
liquid at $T =$ 200\,K.
Although the pair correlation functions appear to be similar for all of
our liquid-like states of water at a wide range of conditions, the
higher-order moments reveal distinct structural differences, e.g.,
related to the degree of tetrahedrality or to how close the density
fluctuations are to Gaussian. While such structural differences can
easily be missed by two-body characteristics, they are clearly captured
by the higher-order moments. Hence, our results further motivate the
need for experimental methods to ascertain three- and higher-body 
correlations in water systems, e.g., via isothermal pressure derivatives
of the structure factor or a spherical harmonic
analysis.\cite{soper_estimating_1995, dhabal_triplet_2014,
dhabal_probing_2017}

A key insight of our higher-order moment analysis is the newly found
temperature non-monotonicity in the Gaussian distance metric
$\fn{l_2}{R}$. First of all, the convergence or non-convergence to a
Gaussian distribution distinguishes the disordered from the ordered
phases. Specifically, in contrast to the two crystalline ice phases, the
distributions of all our disordered states become close to a Gaussian
for large radii as measured by $\fn{l_2}{R}$. For these liquid-like
states, we observe that, at ambient conditions, tetrahedral order and
thermal fluctuations balance out, and hence $l_2(R)$ converges to zero
most rapidly. The convergence slows down at higher and lower
temperatures; in the former case, because of larger thermal
fluctuations; in the latter case, because of higher tetrahedral order.

This non-monotonicity distinguishes itself from the well-known anomaly
in the isothermal compressibility in at least two aspects. Even though
the isothermal compressibility can be related to density fluctuations,
more precisely, the asymptotic number variance, it only holds in the
limit of infinite radii and only pertains to the second moment. In our
case, we find that tetrahedrality also induces non-monotonic temperature
dependencies locally for the higher-order moments. 

Another intriguing aspect of water's anomalies is the nested
structural, dynamic, and thermodynamic anomalies noted by
\citet{errington_relationship_2001}. Specifically, they observe a region
of ``structural'' anomalies, where water's translational and
orientational order decrease upon compression, a region of ``dynamic''
anomalies, where water's diffusion coefficient increases upon
compression, and a region of ``thermodynamic'' anomalies, where density
decreases upon cooling at constant pressure. The set of temperatures and
densities that define the region of structural anomalies completely
contains the region of dynamic anomalies, which in turn contains the
region of thermodynamic anomalies. Given that our present approach
provides a platform to characterize higher-order structural information
about water at all length scales, in future work, it would be
interesting to study how the skewness, kurtosis, and Gaussian distance
metric might behave at state points in the vicinity of water's nested
anomalies. Such an effort may reveal further structural links between
water's anomalous local, mesoscale, and macroscale
phenomena.\cite{errington_relationship_2001}

A consequence of this non-monotonic behavior is that higher-order
moments are no longer negligible relative to the first and second moments
once we consider water states away from ambient conditions. Hence,
second-order approximations that assume Gaussian density fluctuations
and that work well close to room temperature will become less accurate
at both high and low temperatures. We demonstrate this effect for the
information-theoretic approximation of hydrophobic
solubility.\cite{Hu96} With the prominent exception of water at ambient
conditions, we find that the information-theoretic approximation is
accurate only in a range where the constrained moments essentially
determine the entire probability distribution. Therefore, the
higher-order moments generally improve the approximation for the larger
solute radii as one departs from ambient conditions. 

Moreover, for radii beyond the range where the first two moments provide
accurate predictions, the sign of $\gamma_1(R)$ indicates whether the
information-theoretic approximation over- or underestimates the
hydrophobic solubility for our states. The sign of $\gamma_1(R)$, in
turn, is linked to the question whether tetrahedrality or thermal
fluctuations have a greater influence on the local degree of order and
disorder.
Thus, our higher-order moment analysis not only improved the predictions
quantitatively and provides estimates for the range where the
approximation is accurate and which deviations are to be expected, but
the higher-order moments provide us with physical insights into what
determines the local structural features of a wide variety of water
phases. 

One of the most immediate  applications of hydrophobic solubility is in
biology, which of course is typically restricted to near-ambient
conditions. Thus, the utility of the information-theoretic approximation
even with $M=2$ is readily apparent. However, there are many
technologically and scientifically impactful cases where predicting
water's interactions with hydrophobic solutes may also be important at
conditions far from ambient, such as for understanding the impact of
hydrophobic aerosol particles in the
atmosphere\cite{tabazadeh_surface_2002, yang_ice_2021} (i.e.,
low-temperature), or for understanding life in exotic environments such
as on astronomical bodies, e.g., comets, planets,\cite{mottl_water_2007}
(i.e., low-temperature, low-pressure) or near hydrothermal
vents\cite{mcdermott_pathways_2015} (i.e., high-temperature,
high-pressure). In such cases, considering higher-order moments of
$P[N(R)]$ may prove useful.

Finally, we point out that the higher-order moments generically provide
a robust characterization of $n$-body correlations in water's
tetrahedral network. Such additional structural information can, for
example, enhance or complement recent unsupervised machine learning
approaches to classify high- and low-density structures in liquid
water.\cite{donkor_beyond_2024} One could also envision combining these
structural characterization techniques with recent advances in
machine-learned interaction potentials~\cite{behler_perspective_2016,
friederich_machine-learned_2021, wen_deep_2022} to explore structure
across length scales as derived from ab initio models. More broadly, our
findings could also generalize to other highly-polar (e.g., ammonia, HF)
and/or network-forming liquids (e.g., Si, Ge)~\cite{hujo_rise_2011}.

\begin{acknowledgments}
We thank G.\,T.~Barkema and N.~Mousseau for providing their samples of
CRNs. This work was supported in part by the National Science Foundation
Grant No. CBET-2133179 and the U.S. Army Research Office under
Cooperative Agreement No. W911NF-22-2-0103. M.A.K.~acknowledges funding
and support by the Deutsche Forschungsgemeinschaft (DFG, German Research
Foundation) through the SPP 2265, under grant numbers KL 3391/2-2, WI
5527/1-1, and LO 418/25-1, as well as by the Initiative and
Networking Fund of the Helmholtz Association through the Project
``DataMat''. This work used computational resources managed by Princeton
Research Computing, a consortium of groups including the Princeton
Institute for Computational Science and Engineering (PICSciE) and the
Office of Information Technology's High Performance Computing Center and
Visualization Laboratory at Princeton University.
\end{acknowledgments}

\section*{Data availability}
The data that support the findings of this study are available
from the corresponding authors upon reasonable request.

\begin{figure}[h]
  \centering
  \includegraphics[width=0.45\textwidth]{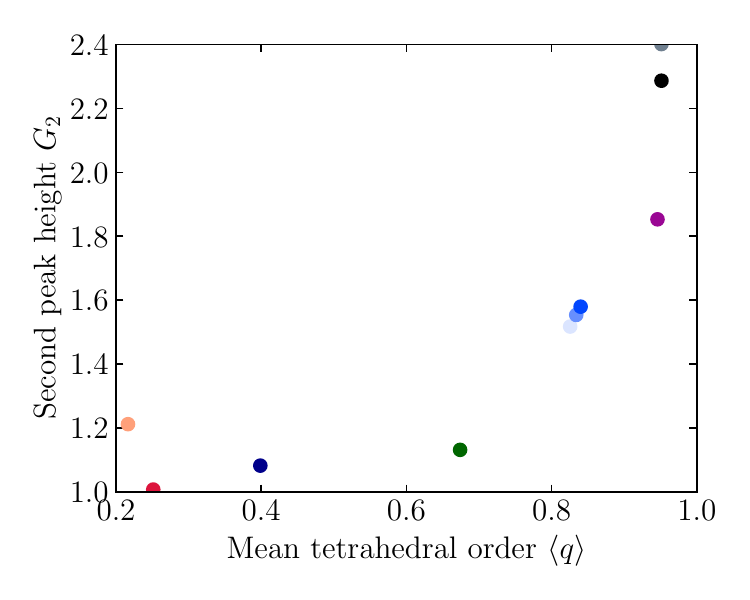}
  % Second-peak height $G_2$
  \caption{The two-body measure $G_2$, which is the height of the
  second peak of $g_2(r)$, is plotted as a function of the mean
  tetrahedral order $\E{q}$ for all states considered here. These states
  are represented by the same color code as in Fig.~\ref{fig:pair-statistics}.
  }
  \label{fig:G2}
\end{figure}

\appendix 

\section{Correlation of the second peak of $g_2(r)$ with tetrahedral order}
\label{app:G2}

The value of the two-body correlation function $g_2(r)$ at the second
peak, which is here denoted by $G_2$, has been employed by
\citet{sellberg_ultrafast_2014} to quantify the degree of tetrahedral
order of liquid water. For purposes of comparison, we apply this
parameter to all states considered here. Figure \ref{fig:G2} clearly
shows a positive correlation between $G_2$ and the mean tetrahedral
order $\E{q}$ for all of the states of water considered in this paper
(and similar to Supplementary Fig.~S21 from
\citet{sellberg_ultrafast_2014} in a different representation for
temperatures 200--340\,K).

\begin{widetext}
\section{Formulas for the skewness and excess kurtosis in terms of $n$-body correlation functions} \label{app:moments-gn}

Here, we represent the expressions for the skewness $\fn{\gamma_1}{R}$
and excess kurtosis $\fn{\gamma_2}{R}$ associated with the spherical
observation window of radius $R$ in terms of the $n$-body correlation
functions $\fn{g_n}{\vect{r}^n}$; see \citet{torquato_local_2021} for
derivations. The skewness, defined in Eq. \eqref{eq:skewness}, can be
written as 
\begin{align}
    \fn{\gamma_1}{R} = [\nv{R}]^{-3/2} 
    \left\{\rho v_1(R) +3\rho^2\int_{\mathbb{R}^d}  h({\bf r}) v_2^{\mbox{\scriptsize int}}(r;R) d{\bf r}+ \rho^3\int_{\mathbb{R}^d}  \int_{\mathbb{R}^d} [g_3({\bf r}^3) -3g_2({\bf r}_{12}) +2] 
v_3^{\mbox{\scriptsize int}}({\bf r}^3;R) d{\bf r}_2 d{\bf r}_3 \right\},   \label{eq:skewness-gn}
\end{align}
where $\fn{v_n^\mathrm{int}}{\vect{r}^n; R}$ is the intersection volume
of $n$ spheres of radius $R$ centered at positions $\vect{r}^n={\bf
r}_1,{\bf r}_2,\cdots,{\bf r}_n$. We see that $\gamma_1(R)$ encodes up
to three-body information. The excess kurtosis, defined in Eq.
\eqref{eq:kurtosis}, can be written as 
\begin{align}
    \fn{\gamma_2}{R} &= [\nv{R}]^{-2} 
    \Bigg\{
    \rho v_1(R) +7\rho^2\int_{\mathbb{R}^d}  h({\bf r}) v_2^{\mbox{\scriptsize int}}(r;R) d{\bf r} + 6\rho^3\int_{\mathbb{R}^d}  \int_{\mathbb{R}^d} [g_3({\bf r}^3) -3g_2({\bf r}_{12})+2] v_3^{\mbox{\scriptsize int}}({\bf r}^3;R) d{\bf r}_2 d{\bf r}_3 \nonumber \\
& + \,\rho^4\int_{\mathbb{R}^d}  \int_{\mathbb{R}^d}  \int_{\mathbb{R}^d}
[g_4({\bf r}^4) -4g_3({\bf r}^3) +12g_2({\bf r}_{12})  -6 ]
v_4^{\mbox{\scriptsize int}}({\bf r}^4;R)  d{\bf r}_2 d{\bf r}_3 d{\bf r}_4  - 3\left[\rho^2 \int_{\mathbb{R}^d}  g_2({\bf r})
v_2^{\mbox{\scriptsize int}}(r;R) d{\bf r}\right]^2 \Bigg\}.    \label{eq:kurtosis-gn}
\end{align}
We see that $\gamma_2(R)$ encodes up to four-body information.

\end{widetext}

%\bibliography{new,paper}

\end{document}